\documentclass[pdflatex,sn-mathphys-num,iicol]{sn-jnl}

\usepackage{graphicx}%
\usepackage{multirow}%
\usepackage{amsmath,amssymb,amsfonts}%
\usepackage{amsthm}%
\usepackage{mathrsfs}%
\usepackage[title]{appendix}%
\usepackage{xcolor}%
\usepackage{textcomp}%
\usepackage{manyfoot}%
\usepackage{booktabs}%
\usepackage{algorithm}%
\usepackage{algorithmicx}%
\usepackage{algpseudocode}%
\usepackage{listings}%
\usepackage{float}  %
\usepackage{comment}  %
\usepackage{booktabs,siunitx,threeparttable}
\theoremstyle{thmstyleone}%
\theoremstyle{thmstyletwo}%
\newtheorem{remark}{Remark}%

\theoremstyle{thmstylethree}%
\newtheorem{definition}{Definition}%

\newcommand\blfootnote[1]{%
  \begingroup
  \renewcommand\thefootnote{}\footnote{#1}%
  \addtocounter{footnote}{-1}%
  \endgroup
}
\begin{document}

\title[Article Title]{
Reconstructing Brain Causal Dynamics for Subject and Task Fingerprints using fMRI Time-series
Data \vspace{-0.5em}}







\author[1]{\fnm{Dachuan} \sur{Song}}

\author[2]{\fnm{Li} \sur{Shen}}

\author[2]{\fnm{Duy} \sur{Duong-Tran}}

\author*[1]{\fnm{Xuan} \sur{Wang}}\email{xwang64@gmu.edu}

\affil*[1]{\orgdiv{Department of Electrical and Computer Engineering}, \orgname{George Mason University}, \orgaddress{\city{Fairfax}, \state{Virginia}, \country{USA}}}

\affil[2]{\orgdiv{Department of Biostatistics, Epidemiology and Informatics}, \orgname{Perelman School of Medicine, University of Pennsylvania}, \orgaddress{\city{Philadelphia}, \state{Pennsylvania}, \country{USA}}}

\abstract{
\textbf{Purpose}: Recently, there has been a revived interest in system neuroscience causation models, driven by their unique capability to unravel complex relationships in multi-scale brain networks.
In this paper, we present a novel method that leverages causal dynamics to achieve effective fMRI-based subject and task fingerprinting.
\textbf{Methods}: By applying an implicit-explicit discretization scheme, we develop a two-timescale linear state-space model. Through data-driven identification of its parameters, the model captures causal signatures, including directed interactions among brain regions from a spatial perspective, and disentangled fast and slow dynamic modes of brain activity from a temporal perspective. 
These causal signatures are then integrated with: (i) a modal decomposition and projection method for model-based subject identification, and (ii) a Graph Neural Network (GNN) framework for learning-based task classification. 
Furthermore, we introduce the concept of the brain reachability landscape as a novel visualization tool, which quantitatively characterizes the maximum possible activation levels of brain regions under various fMRI tasks. 
\textbf{Results}: We evaluate the proposed approach using the Human Connectome Project dataset and demonstrate its advantage over non-causality-based methods. The obtained causal signatures are visualized and demonstrate clear biological relevance with established understandings of brain function.
\textbf{Conclusion}:  We verified the feasibility and effectiveness of utilizing brain causal signatures for subject and task fingerprinting. Additionally, our work paves the way for further studies on causal fingerprints with potential applications in both healthy controls and neurodegenerative diseases.}

\keywords{fMRI fingerprinting, Brain causal dynamics}



\maketitle

\section{Introduction}\label{sec1}
\blfootnote{A preliminary version of this work appeared as~\cite{SDC-SL-DTD-WX-2024} at the 2024 ACM-BCB Conference.}
Advancements in brain imaging technologies, especially Functional Magnetic Resonance Imaging (fMRI) of Blood-Oxygen-Level-Dependent (BOLD) signals, have enabled precise and quantitative measures of brain activity, leading to a significant boost in neuroscience research~\cite{logothetis2002neural}. Among various brain modeling approaches, causation models have recently received a revamped interest~\cite{kiebel2007dynamic,ross2024causation} in recent years, due to their capability to reveal complex relationships without multi-scale brain networks. These models have been extended in multiple directions to understand the brain's underlying processes~\cite{mannino2015foundational}, explaining how these processes support cognition and behavior~\cite{rigoux2015dynamic,duong2021morphospace,garai2023mining}, and propose new methods to regulate brain activity~\cite{gu2015controllability,amico2021toward} for enhancing normal functions and mitigating disorders~\cite{xu2022consistency}.
However, most fingerprinting pipelines still rely on static, symmetric connectivity maps and thus miss the directed, multi-timescale dynamics that underlie neural communication.

Inspired by these developments, this paper explores the use of causal dynamic models to extract informative features from the brain for fMRI fingerprinting. Specifically, we aim to leverage causal signatures derived from fMRI time-series data to identify both the individual subject (subject fingerprinting) and the cognitive task being performed (task fingerprinting). A major challenge in this process arises from the variability across subjects and tasks, along with the limited number of recordings available for each subject-task combination~\cite{SM-SM-VI-DI-BE}.
In other words, the problem combines heterogeneity (across subjects/tasks) with data sparsity, which calls for models that are both interpretable and sample-efficient.
Our approach focuses on uncovering and utilizing latent causal signatures in the brain. While this requires more specialized model structures, it can offer improved data efficiency. This distinguishes our method from traditional correlation-based approaches~\cite{abbas2020geff,abbas2023tangent,amico2021toward,duong2021morphospace,chiem2022improving,duong2024homological,duong2024principled,FES-SX-SD-RMD-HJ:15,amico2018quest}, which are limited in their ability to capture the directional, evolving cause-and-effect relationships that are fundamental to cognitive dynamics.

\textit{Statement of contribution:} In this paper, our result provides a solid `YES' answer to the feasibility and effectiveness of utilizing brain causal signatures for subject and task fingerprinting. Our main contributions are summarized as follows:

\begin{itemize}
    \item Introducing a two-timescale linear state-space model, which (i) captures directed interactions among brain regions from a spatial perspective, and (ii) disentangles fast and slow dynamic modes of brain activity from a temporal perspective. Model parameters are identified using a data-driven, implicit-explicit discretization scheme.

    \item Integrating the causal signatures with (i) a modal decomposition and projection method for model-based subject identification, and (ii) a Graph Neural Network (GNN) framework for learning-based task classification.

    \item Leveraging the control-theoretic interpretation of the causal-based model to introduce the concept of Reachability landscape as a novel visualization tool, which quantitatively characterizes the maximum possible activation levels of brain regions under various fMRI tasks. 
\end{itemize}
Extensive experiments and comparative analyses are conducted to validate the proposed approach. This research also lays the groundwork for future exploration of causal fingerprinting in both healthy individuals and those affected by neurodegenerative disorders.

\section{Related work}
The concept of an fMRI “fingerprint”~\cite{abbas2020geff,abbas2023tangent,amico2021toward,duong2021morphospace,chiem2022improving,duong2024homological,duong2024principled,duong2024preserving,duong2024principled_m,nguyen2024volume} can be considered as a classification problem (identifying which subject or task corresponds to a given fMRI sample). 
In conventional applications, data richness has a major impact on classification accuracy. However, in data-scarce settings, model structure and algorithmic interpretability become critical for maintaining robust performance. For example, in visual person re-identification (re-ID) tasks, incorporating auxiliary features such as gender, color, and texture can significantly improve identification accuracy~\cite{YM-SJ-LG-XT-SL:21}. However, a similar idea is not directly applicable to fMRI-based applications, as raw brain signals are inherently difficult to interpret or to translate into distinguishable features.

\subsection{Static connectivity models}
In neuroscience research, one strategy to address this challenge is to use the functional connectome (FC)~\cite{SM-SM-VI-DI-BE}. The FC measures the correlations of the fMRI time-series data among pair-wise brain regions during rest or task conditions, and it has been successfully utilized as individual-specific neural signatures to classify various cognitive states and neurological disorders~\cite{CB-ZG-ZA-XL-HW-SJ-WT-CV-WY:21,FES-SX-SD-RMD-HJ:15}. However, FC is static and undirected, thus missing causal directionality. Another common brain-network modeling technique is Independent Component Analysis (ICA). This approach separates fMRI data into spatially distinct, statistically independent components corresponding to different brain networks~\cite{CA-VD-AD-TU-PE,MC-MJ-MA-SC-BR}. It has proven especially useful for revealing resting-state networks and explaining their roles in cognitive function and pathology~\cite{BE-CF-DE-MA-DE,SM-SM-FO-PT-MI:2009,VD-MP-PO-HE:2010}.
Although FC and ICA have achieved notable success in many applications, they still have several limitations. In particular, they do not fully capture the dynamic nature of brain activity, as they lack the temporal resolution needed to detail interactions among regions, and cannot represent the directionality of connections between regions~\cite{CB-ZG-ZA-XL-HW-SJ-WT-CV-WY:21,UM-SL-DD-WX:23,abbas2023tangent,duong2024homological}. Each of these aspects, dynamics, temporal detail, and directionality, is essential for understanding the cause-and-effect interactions within brain networks.

\subsection{Causal connectivity models}
In contrast to static correlational methods, causality-based approaches such as Dynamic Causal Modeling (DCM) and Granger causality offer alternative insights into patterns of brain interactions. However, these directed connectivity methods have been relatively under-explored in the context of fMRI fingerprinting. Analyzing causality in fMRI entails examining directional influences among brain regions over time, which provides a window into the brain’s dynamic modes of interaction. For example, DCM uses a Bayesian framework to model interactions among latent neural states with bilinear differential equations, which describes how activity in one region influences others over time~\cite{FR-KJ-HA-L-P-W,S-KE-P-WD-D}. Similarly, Granger causality evaluates whether one time series can predict another, thereby identifying potential directed relationships between neural signals~\cite{GR-CW,B-AB-B-L,R-A-F-E-G}. 
Yet such directed methods remain under-explored for fMRI fingerprinting. 
Most Dynamic Causal Modeling (DCM) or Granger-causality studies restrict the analysis to a small number of regions of interest (ROIs) and assume multiple sessions per subject. 
Consequently, it is unclear how well these approaches scale to whole-cortex parcellations with few runs per subject, or to joint subject and task fingerprinting. 
To further enhance these causal models and better capture complex neural dynamics, recent research has introduced several modifications.
For instance, threshold-based structures have been incorporated to reflect the saturation of neural activation~\cite{wang2021data,wang2023efficient,wang2022data}, and multifactorial dynamics have been proposed to represent systems that operate across multiple time scales~\cite{iturria2017multifactorial}.

\subsection{Learning-based approaches}
Beyond these correlation-based or causality-based methodologies, classical machine-learning algorithms also remain popular. Random Forest (RF)~\cite{BR-LE-2001} and Support Vector Machines (SVM)~\cite{HM-DS-OE-PJ-SB-1998} provide high interpretability and easy deployment across diverse datasets.
Recent studies on fMRI fingerprinting have also explored a variety of advanced techniques to capture different facets of brain dynamics and connectivity patterns. These include deep-learning architectures. Deep Neural Networks (DNN) capture high-dimensional connectivity patterns~\cite{WX-LX-JZ-NB-ZY-2020}. Deep Bidirectional Recurrent Neural Networks (DBRNN) model temporal dynamics in brain signals~\cite{ZH-SH-FA-LO-WU:2022}. Brain Attend and Decode (BAnD) applies attention mechanisms to highlight salient brain regions~\cite{nguyen2020attend}.
A recent variant, TCN--BiLSTM, combines dilated Temporal Convolutional Networks (TCNs) with bidirectional Long Short-Term Memory (BiLSTM) layers to model short- and long-range dependencies in fMRI sequences~\cite{KM-GJ-BJ-BK-2025}. However, it still operates on undirected regional time series and lacks an explicit multi-timescale causal framework, which limits its ability to characterize directed network interactions.
Geometry- and topology-based pipelines have also been proposed. Tangent-Space Fingerprinting (TSF) maps functional-connectivity matrices to the Riemannian tangent space of the symmetric positive-definite (SPD) manifold, capturing global correlation geometry while still treating connectivity as static and undirected~\cite{CD-KV-DJ-2025}. The PH-H0 landscape describes networks via zero-dimensional persistent homology (H0), providing a noise-robust topological summary yet discarding temporal scale and edge direction~\cite{WY-XJ-CY-YY-2025}.
Classical classifiers such as Random Forests and Support Vector Machines are trained on pre-specified connectivity vectors. Without directionality or temporal context, their accuracy soon plateaus and lags behind causal models~\cite{BR-LE-2001,HM-DS-OE-PJ-SB-1998}.
End-to-end deep networks (DNN, DBRNN, BAnD) can harvest richer patterns, yet they require large training sets and remain hard to interpret.  
Geometry-based TSF and topology-based PH-H0 offer elegant dimensionality reduction, but both treat connectivity as static and undirected~\cite{WY-XJ-CY-YY-2025}.  
Across these lines of work, directed interactions and the multiple time scales that shape cognition are largely ignored.  
Our method fills this gap: a single scan yields compact causal signatures that encode directionality and tempo, and a lightweight GNN turns them into accurate fingerprints while preserving biophysical meaning.

\section{Methods: Causal Fingerprint}



In this paper, the concept of causal fingerprint is defined as follows: {(later referred to as fingerprint, for simplicity.)}
\begin{definition}
    Causal fingerprint is the degree to which a subject or an fMRI task can be identified from a labeled database, particularly based on the subjects' or fMRI tasks' unique \textit{cause-and-effect} cognitive signatures.
\end{definition}

To summarize our method, we introduce an implicit-explicit discretization method that yields a two-timescale state-space model to capture causal signatures. Then we combine the causal signatures with a modal decomposition and projection method for subject fingerprinting and with a GNN model for task fingerprinting. Finally, building on the state-space model, we propose a new visualization tool that characterizes the reachability of the brain state, which quantitatively represents the maximally possible excitation level for different brain regions.

\noindent \textbf{Notations.} Throughout this section, suppose we are given an fMRI data set with $\mathcal{S}$ representing the set of all subjects, and $\mathcal{H}$ representing the set of all tasks performed by the subjects. 
Let $d_{s,h}\in\mathbb{R}^{p\times T}$ represent the time-series recording of a subject $s\in\mathcal{S}$ performing a fMRI task $h\in\mathcal{H}$. 
The dimension $p$ is the number of brain parcellations and $T$ is the testing duration. Let $d_{s,h}(k)\in\mathbb{R}^{p\times T}$ represent the $k^{th}$ column of $d_{s,h}$. Let $\text{vec}(\cdot)$ represent the vectorization (stacking all columns) of a matrix into a single-column vector.

\subsection{A Two-timescale State-space Model for Causal Dynamics}\label{sec_model}
\begin{figure}[t]
\centering
\includegraphics[width=.45\textwidth]{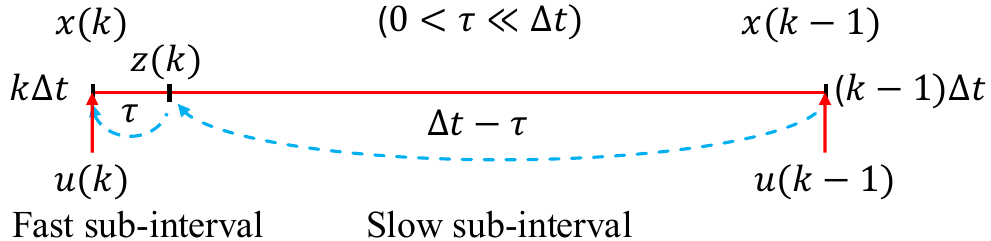}
\caption{Two–timescale partition of a single fMRI sampling period $[(k-1)\Delta t,\,k\Delta t]$.  
The interval is split into a fast sub‑interval of duration $\tau$ (short solid red) and a slow sub‑interval of duration $\Delta t-\tau$ (long solid red), with $0<\tau\ll\Delta t$.  
Driven by the slower input $u(k-1)$ and the current input $u(k)$, the state evolves from the previous scan $x(k-1)$, passes the intermediate point $z(k)=x(k-\tau)$, and reaches the current scan $x(k)$.}
\label{fig_timescale}
\end{figure}

\begin{figure*}[t]
\centering
\includegraphics[width=.8\textwidth]{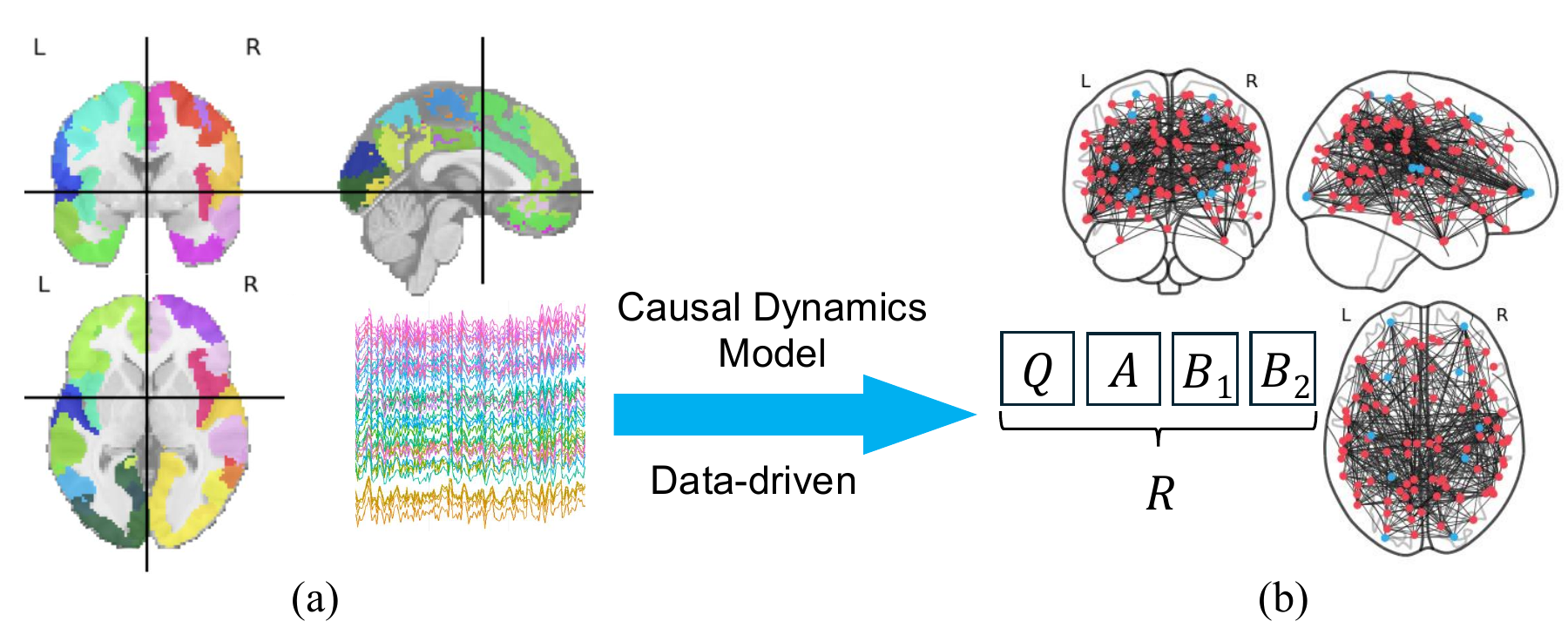} 
\caption{\textbf{(a)} Brain Activity (time-series data) captured using Schaefer Parcellation. \textbf{(b)} Causal Dynamics modeling and data-driven parameter identification.}
\label{fig:brain_connect}
\end{figure*}

To capture the variability of subjects performing different fMRI tasks, we propose a two-time-scale linear state-space model that characterizes the brain’s activity both spatially and temporally. Spatially, the model captures directed interactions among brain regions, identifying which areas causally influence others. Temporally, it disentangles brain dynamics across two time-scales, characterizing both rapid neural signaling and slower hemodynamic responses. 

Although fMRI only records data at a low sampling rate, the brain operates on multiple timescales~\cite{PO-JR-LE-LD-2021}:
fast neuronal activities such as synaptic firing and local synchrony occur on the order of milliseconds; while slow hemodynamic responses unfold over several seconds. As a result, the observed fMRI time-series data ($d_{s,h}$) represents an aggregation of these intertwined processes, making it challenging to reconstruct the parameters of the underlying dynamics. To address this, we propose an implicit-explicit discretization method that partitions each sampling period into a fast and a slow sub-interval. The fast dynamics capture concurrent interactions that reflect rapid neuronal activity, while the slow dynamics capture hemodynamic responses and long-timescale network interactions among brain regions.

In the following, we first present the proposed two-time-scale model, then we validate the model by showing how it is mathematically derived and how the model parameters can be obtained through a data-driven method. Since this modeling framework applies generically to any subject and any task, we omit the subscripts $s,h$ for clarity in what follows.

Consider fMRI data with $p$ brain parcellations. We divide the parcellations into two subsets: system states $x(k)\in\mathbb{R}^m$ and inputs $u(k)\in\mathbb{R}^n$, with $m+n=p$. Each entry of $x(k)$ or $u(k)$ represents the activity of a specific brain region at discrete time $k$. We assume that $k\in\mathbb{Z}$ indexes successive fMRI scans, typically spaced $\Delta t$ seconds apart (e.g. $\Delta t=0.72s$ for the dataset used in this paper). Our aim is to construct a two-timescale state-space model of the following form:
\begin{align}\label{eq_model}
    x(k) = Qx(k) \!+\! Ax(k-1) \!+\! B_{1}u(k)\!+\! B_{2}u(k-1).
\end{align}
Here, $Q \in \mathbb{R}^{m \times m}$ captures the fast sub-interval (concurrent) interaction among the neuronal states; $A \in \mathbb{R}^{m \times m}$ encodes the slow sub-interval (cross-lagged) transition from $x(k-1)$; $B_{1}\,u(k)$ and $B_{2}\,u(k-1)$ incorporate the inputs into the system for fast and slow dynamics, respectively. 

\medskip
\noindent\textbf{Implicit-Explicit discretization:}
To systematically derive the two-timescale discrete model~\eqref{eq_model}, we divide each fMRI sampling period into slow and fast sub-intervals, as illustrated in Fig.~\ref{fig_timescale}. Consider a continuous-time dynamical system with both slow and fast modes:
\begin{align}
    \frac{d}{dt} x(t)=F_\mathrm{s}x(t)+G_\mathrm{s} u(t)+ \frac{1}{\epsilon}(F_\mathrm{f} x(t)+G_\mathrm{f} u(t)),
\end{align}
where $0<\epsilon \ll 1$ indicates the seperation of timescales. To discretize the system, we divide the full sampling interval $\Delta t$ into a slow component $\Delta t - \tau$ and a fast component $\tau$, such that the ratio of their lengths satisfies: $\displaystyle \frac{\Delta t - \tau}{\tau}=\frac{1}{\epsilon}$ \footnote{In practice, $\tau$ is typically on the order of 0.002s~\cite{KA-ER-SC-JH-JE-2000}, thus $\frac{1}{\epsilon}\approx\frac{1}{360}$.}.

The slow dynamics are approximated using an explicit Euler approximation from $x(k-1)$ to an intermediate state $z(k)$:
\begin{align}
    \frac{z(k)-x(k-1)}{\Delta t - \tau}=\Bigl[F_{\mathrm{s}}\,x(k-1) + G_{\mathrm{s}}\,u(t-1)\Bigr].
\end{align}
which yields
\begin{align}\label{eq_updatez} 
z(k) \!=\! x(k\!-\!1) \!+\! (\Delta t \!-\! \tau)\left[F_{\mathrm{s}}x(k\!-\!1) \!+\! G_{\mathrm{s}}u(k\!-\!1)\right]
\end{align}

The fast dynamics are then modeled as an update over the interval $\tau$, which starts from $z(k)$ and evolves to $x(k)$:
\begin{align}\label{eq_impli}
    \frac{x(k)-z(k)}{ \tau}=\Bigl[F_\mathrm{f} x(k)+G_\mathrm{f} u(k)\Bigr].
\end{align}
which yields: 
\begin{align}\label{eq_updatefast}
    x(k)=\tau\Bigl[F_\mathrm{f} x(k)+G_\mathrm{f} u(k)\Bigr]+ z(k).
\end{align}
Note that the intermediate state $z(k)$ is not directly observable in the fMRI time series, which only samples the system at the slower rate $\Delta t$. Thus, we introduce an implicit Euler method in~\eqref{eq_impli} that uses $x(k)$ and $u(k)$ to approximate the update. This allows us to cancel out the intermediate state  $z(k)$ and use the one-time scale fMRI time-series data to reconstruct the parameters of the two-time-scale systems. Specifically, by incorporating~\eqref{eq_updatez} and~\eqref{eq_updatefast}, one has
\begin{align}
     x(k)=&~\tau F_\mathrm{f} x(k) + \bigl[I + (\Delta t - \tau)\,F_{\mathrm{s}}\bigr]\,x(t{-}1)\nonumber\\
     & + \tau G_\mathrm{f} u(k) + (\Delta t - \tau)G_{\mathrm{s}} u(k-1).
\end{align}

Now, define 
\begin{align*}
Q &\triangleq \tau\,F_{\mathrm{f}}, 
\quad 
A \triangleq I + (\Delta t - \tau)\,F_{\mathrm{s}}, \\
B_{1} &\triangleq \tau\,G_{\mathrm{f}},
\quad 
B_{2} \triangleq (\Delta t - \tau)\,G_{\mathrm{s}}.
\end{align*}
The update equation aligns with model~\eqref{eq_model}.

\medskip
\noindent\textbf{Data-driven reconstruction of model parameters:}
Given fMRI data $\{x(k),u(k)\}$ for $k=0,\dots,T$, each time $k\ge1$ should satisfy equation~\eqref{eq_model}.  
Define the data matrices $X^{1:T}=\begin{bmatrix}
    x(1)& x(2)&\cdots x(T)
\end{bmatrix}$ and $U^{0:T-1}=\begin{bmatrix}
    u(0)& u(1)&\cdots u(T-1)
\end{bmatrix}$. 
A compact representation of the dynamics follows
$$X^{1:T}=QX^{1:T} + AX^{0:T-1} + B_1U^{0:T-1}+ B_2U^{1:T}$$
To determine the model parameters in the presence of model imperfections and measurement noise, we formulate the generalized least squares problem:
\begin{align}\label{eq_matrix_id}
       &\underset{Q, A,B_1,B_2}{\arg\min}~   \lambda(\|Q\|_F+\|A\|_F+\|B_1\|_F+\|B_2\|_F) \nonumber\\ 
       & \!+\! \big\|(Q\!-\!I)X^{1:T} \!\!+\! AX^{0:T\!-\!1}\!\!+\! B_1U^{0:T\!-\!1}\!\!+\! B_2U^{1:T}\big\|_F \nonumber\\
     & \text{subject to}\quad Q_{ii}=0,~\forall i\in\mathbf{m} 
\end{align}

Here $Q,A\in\mathbb{R}^{m\times m}$, $B_{1},B_{2}\in\mathbb{R}^{m\times n}$,  
$X^{1:T},X^{0:T-1}\in\mathbb{R}^{m\times T}$, and  
$U^{0:T-1}, U^{1:T}\in\mathbb{R}^{n\times T}$,  
where $\lambda>0$ controls the strength of the regularization to prevent over-fitting, and $\|\cdot\|_{F}$ denotes the Frobenius norm (for any matrix $M\!\in\!\mathbb{R}^{p\times q}$, $\|M\|_{F}=\sqrt{\sum_{i=1}^{p}\sum_{j=1}^{q}M_{ij}^{2}}$)~\cite{peng2016connections}. 
The residual term  
$\bigl\|(Q-I)X^{1:T}+AX^{0:T-1}+B_{1}U^{0:T-1}+B_{2}U^{1:T}\bigr\|_{F}$  
is the Frobenius norm of an $m\times T$ matrix, which aims to extract system parameters as causal signatures by minimizing the difference between the system's measured next time state with the next time state predicted by our two-time scale model. The  
$\|Q\|_{F}$, $\|A\|_{F}$, $\|B_{1}\|_{F}$ and $\|B_{2}\|_{F}$ regularization terms to prevent over-fitting.
By solving~\eqref{eq_matrix_id}, we transform a complex fMRI time-series data $d(t)=\{x(t), u(t)\}$ into a structured and meaningful representation ${R}= [Q~A~B_1~B_2]\in\mathbb{R}^{m\times(2m+2n)}$ of brain dynamics signature, as visualized in Fig.~\ref{fig:brain_connect} .
This transformation not only makes the data more accessible for subsequent analysis but also reveals the patterns of `spatio-temporal’ causality relationships that underpin cognitive processes.
In the following, we will build on $R$ to perform subject and task fingerprints.

\begin{remark}\label{RM_1} [\textbf{\textit{Model choice:}}]
In contrast to traditional functional connectome (FC) representations used for brain fingerprinting, the proposed causal signature offers two key advancements: it captures directional interactions (i.e., which region inhibits or excites another) and temporal dependencies (i.e., how earlier neural activities influence subsequent ones across two time-scales). In comparison, the FC method only captures undirected and concurrent relationships between the activities of pairwise brain regions.
From a modeling perspective, we note that this work adopts a linear state-space model~\eqref{eq_model}, which represents a relatively simple form of causal modeling compared to the broader literature that incorporates more complex, nonlinear structures—such as bilinear dynamics~\cite{FR-KJ-HA-L-P-W}, threshold dynamics~\cite{wang2021data}, and multifactorial time-varying dynamics~\cite{iturria2017multifactorial}.
However, we argue that the two-timescale structure enabled by implicit-explicit discretization already encodes sufficiently rich features. The use of a simpler model (linear v.s. nonlinear) does not diminish the results of the paper; instead, it provides compelling evidence for the utility of cause-and-effect signatures in fingerprinting applications. Moreover, simpler models generally have mild requirements on data richness. As will be demonstrated in experiments, the proposed approach achieves accuracy comparable to more complex, non-causal methods (e.g., FC and end-to-end learning), even when using low-resolution fMRI data (Schaefer-100 parcellation).
\end{remark}

\subsection{Subject Causal Fingerprint via Modal Decomposition and Projection}\label{sec_subjetid}

Based on the model~\eqref{eq_model} and the parameter reconstruction~\eqref{eq_matrix_id}, we can obtain a representation $R_{s,h}=[Q~A~B_1~B_2]_{s,h}$ for any subject-task pair that extracts its spatial and temporal signatures across two time-scales. In the following, we incorporate these signatures with a state-space modal decomposition~\cite{dahleh2004lectures} and projection method to perform subject fingerprinting. 

For each fMRI task $h \in \mathcal{H}$, we construct a labeled reference set:
$$\widehat{\mathcal{R}}_h=\{R_{s,h}^D\}, \quad s\in\mathcal{S}.$$ 
where the superscript $D$ denotes data samples with known subject identities. 
We assume that the dataset includes at least two recordings for every subject-task pair. This allows one sample to be used in the reference set $\widehat{\mathcal{R}}_h$, while the remaining samples serve as query instances for evaluation.
Given a query sample with causal signature $R_{h}^Q$ of an unknown identity, the objective is to identify the subject label $s$ by comparing it against the labeled reference set $R_{s,h}^D$.

The key challenge for subject causal fingerprinting lies in the limited data for each subject, as the fMRI data collection is time-consuming.
Essentially, we need to solve a one-shot classification problem~\cite{YM-SJ-LG-XT-SL:21}, i.e., only one sample per subject is available to determine their identity from a large number of candidates.
To address this, we leverage the control-theoretic interpretation of the model~\eqref{eq_model}, in which the causal system signature $R = [Q~A~B_1~B_2]$ naturally encodes the system's \textit{dynamic modes}~\cite[Chapter 12]{dahleh2004lectures}, reflecting invariant subspaces of brain dynamics. These dynamic modes serve as augmented features for robust subject identification.

We extract dynamic modes through eigendecomposition and coordinate transformation by applying them, respectively, to the fast and slow timescales. For example, to analyze the slow dynamic mode, we treat the system as 
\begin{align}\label{eq_perturb}
    x(k) =  Ax(k-1) +  B_{2}u(k-1) + o(k).
\end{align}
where $o(k)=Qx(k)+ B_{1}u(k)$ is considered as a perturbation term. To perform modal decomposition, we express $A=\widehat{T}^{-1}\widehat{\Lambda} ~\widehat{T}$, where $\widehat{\Lambda}$ is the diagonal matrix (or a Jordan matrix if not diagonalizable~\cite{horn2012matrix}) corresponding to the eignevalues of $A$.

Left multiplying $\widehat{T}^{-1}$ to~\eqref{eq_perturb} yields
\begin{align}\label{eq_model3}
    \widehat{x}(t) = \widehat\Lambda \widehat{x}(t-1) + \widehat{B}_2u(t-1) +  \widehat{o}(k),
\end{align}
where $\widehat{x}(t)=\widehat{T}^{-1}{x}(t)$,  $\widehat{B}=\widehat{T}^{-1}B$, and $\widehat{o}(k)=\widehat{T}^{-1} o(k)$. Since $\widehat{\Lambda}$ is diagonal, each state in~\eqref{eq_model3} evolves independently along its corresponding dynamic mode embedded in $\widehat{T}$. 
Similarly, for the fast dynamics, we compute $\bar\Lambda$ and $\bar{T}$ such that $Q=\bar{T}^{-1}\bar{\Lambda} ~\bar{T}$. These capture the concurrent interactions among brain regions by treating the slow dynamics as perturbations.

We use transformation matrices $\widehat{T}$ and $\bar{T}$ as augmented features for subjects to perform fingerprinting, since they encode the dynamic modes of~\eqref{eq_model}. Specifically, we determine the identity of a query sample by minimizing an alignment-based distance:
\begin{align}\label{eq_dist}
\underset{s}{\arg\min}\quad\text{Dist}(\mathcal{T}_{R_{h}^Q},~\mathcal{T}_{R_{s,h}^D})
\end{align}
where $\mathcal{T}$ is a feature set composing $[\widehat{T},~\bar{T}]\in\mathbb{R}^{m\times 2m}$, or a subset of these features, depending on which features best capture inter-subject variation. A detailed discussion on this will be given in the \textit{Results} section.
The distance metric  $\text{Dist}(\cdot,\cdot)$ is computed as the permutation-aligned pairwise vector distance, which compares two sets of vectors, where the elements are unordered. The key idea is to compute the cosine distance between elements in the two sets after finding an optimal permutation alignment that minimize overall distances, specifically, if defining $\mathcal{T}_{R_{h}^Q}=[x_1,\cdots,x_{2m}],~\mathcal{T}_{R_{s,h}^D}=[y_1,\cdots,y_{2m}]$, 
\[
\text{Dist}(\mathcal{T}_{R_{h}^Q},~\mathcal{T}_{R_{s,h}^D})=\min_{\pi \in \mathcal{Z} } \sum_{i=1}^n \left(1- \frac{x_i^{\top} y_{\pi(i)}}{\|x_i\| \|y_{\pi(i)}\|}\right)
\]
where \( \mathcal{Z} \) denotes the set of all permutations of \( \{1, 2, \ldots, 2m\} \).
Note that the eigenvalue matrices $\bar\Lambda$ and $\widehat\Lambda$ quantify the significance of each mode in the dynamics. We, however, do not incorporate them directly in the fingerprinting task. 
Instead, our method treats all dynamic modes as equally informative, regardless of their eigenvalues. This contrasts with dimensionality-reduction methods like Principal Component Analysis (PCA), which retain only the most significant components. Our method, via~\eqref{eq_dist}, operates in the full space spanned by all dynamic modes, under the assumption that when different subjects perform the same fMRI task, their major dynamic modes might be similar but vary in certain minor dynamic modes.

\subsection{Task Causal Fingerprint via Graph Neural Network}
\begin{figure*}[t]
\centering
\includegraphics[width=.6\textwidth]{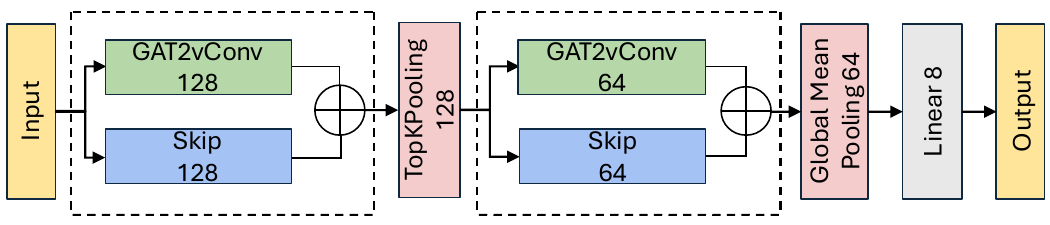} 
\caption{Architecture of the five-layer GNN used for task fingerprinting.}
\label{fig:model_architecture}
\end{figure*}

Unlike subject fingerprint, where the primary challenge arises from the large number of subjects and the limited data per subject, the goal of task fingerprint is to identify a task from a finite set of candidates. In this case, the data is richer, as each task is performed by all participating subjects. Nevertheless, task fingerprinting still faces intrinsic challenges due to inter-subject variability in brain structure and functional performance. As a result, methods relying solely on predefined algebraic operators, as discussed in Section~\ref{sec_subjetid}, are no longer sufficient.

In this subsection, we implement a causal-based task fingerprinting by leveraging the extracted signatures $R_{s,h}$ from Section~\ref{sec_model}. These signatures provide a graph-theoretic representation of brain networks, which we incorporate into a Graph Neural Network (GNN)-based learning model.
To explain our method, we construct a labeled database: 
$$\widehat{\mathcal{R}}=\{R^{D}_{s,h}\}, \quad s\in\mathcal{S}, \text{ and } h\in\mathcal{H}$$ 
where $\mathcal{S}$ and $\mathcal{H}$ are subject sets and fMRI task sets, respectively. 
Given a query $R^Q$ data with unknown identity and task, our goal is to determine the task index $h$ of the query using the database $\widehat{\mathcal{R}}$.

A key advantage of Graph Neural Networks (GNNs) is their ability to model input data as graph structures~\cite{khemani2024review}. This aligns naturally with our data representation, where the matrix $R = [Q~A~B_{1}~B_{2}]$ encodes causal interactions between brain regions. Each row of $R$ captures how a specific brain region is influenced by other regions and external inputs.
We define the GNN model $\mathcal{G}(\mathcal{V}, \mathcal{E})$, where $\mathcal{V}$ represents the set of nodes with $|\mathcal{V}| = m$. The feature of each node is a vector in $\mathbb{R}^{(2m+2n)}$, corresponding to each row of $R$. The normalized matrix $A$ in $R$ is used as the adjacency matrix to initialize the edge features $\mathcal{E}$.

Building on this input structure, we employ a five-layer Graph Neural Network architecture tailored for task fingerprinting from fMRI data.  The network starts with a GATv2Conv layer that re-weights each node’s neighbours, emphasising task-relevant connections~\cite{velickovic2017graph} (input $200$, out $32$ features with $4$ attention heads, dropout $0.30$).  
A parallel linear skip path (output $128$ channels) is summed with the convolution result so the original signal remains visible to deeper layers, improving gradient flow and protecting against over-smoothing~\cite{xu2021optimization}.  TopKPooling then selectively retains the most informative nodes (ratio $0.8$), reducing graph complexity and focusing computation on salient sub-structures.  This sparsification step improves both efficiency and performance.  
A second GATv2Conv block, again accompanied by a skip connection, further refines the retained features and maps them into a more discriminative space for task recognition (input $128$, out $32$ features with $2$ heads, dropout $0.30$; $32\times2=64$ output channels; skip output $64$ channels).  Global-mean pooling aggregates these refined node features across the graph, producing a single $64$-dimensional vector.  A linear head of size $64\times8$, followed by log-softmax, converts this vector into task probabilities for training and evaluation.  During training, we randomly drop $10\%$ of edges to enhance robustness.  The overall model structure is visualised in Figure~\ref{fig:model_architecture}.
The model is trained using supervised learning with data batches drawn from $\widehat{\mathcal{R}}$.  The rows of $R^D_{s,h}$ serve as the input features for each node, while $h$  is used as the corresponding fMRI task label. Detailed training parameters are provided in the Appendix.


\subsection{New Visualization: Brain Reachability Landscape}
\label{sec:reachability}
In control theory, reachability refers to the ability of a control system to move from one state to another using admissible control inputs. In this subsection, we apply this concept to the causal dynamics in~\eqref{eq_model}, which enables a novel visualization tool that characterizes the reachability of the brain state $x(t)$. This generates a quantitative landscape of regional brain excitation level, which complements traditional connectivity measures that focus on the interaction patterns among brain regions.

To achieve this, we first rewrite the model~\eqref{eq_model} into an equivalent state evolution form:
\begin{align}\label{eq_model_ev}
    x(t) & = f(x(t-1),u(t),u(t-1))\nonumber\\
    &= \widehat{A}\,x(t-1) + \widehat{B}_{1}\,u(t)+ \widehat{B}_{2}\,u(t-1),
\end{align}
where
\begin{align*}
\widehat{A}   &\triangleq (I - Q)^{-1}A, \\
\widehat{B}_2 &\triangleq (I - Q)^{-1}B_2, \\
\widehat{B}_1 &\triangleq (I - Q)^{-1}B_1.
\end{align*}
Compared to the original model~\eqref{eq_model}, which explicitly represents the fast-slow interactions in brain dynamics, the reformulated model~\eqref{eq_model_ev} integrates fast interactions directly into the state evolution. This facilitates a direct interpretation of how the system state evolves over time in response to control inputs.

Given a terminal time $T_M$, bounded energy on the control input sequence $U^{0:T_M-1}=~\begin{bmatrix}
    u(0)& u(1)&\cdots u(T_M-1)
\end{bmatrix}$, and zero initial state,  we define the fMRI reachability set as:
\begin{align*}
&\quad \mathcal{B}\triangleq \bigcup x(T_M)\\
s.t. &\quad x(t) = f(x(t-1),u(t),u(t-1)),\\
&\quad \|U^{0:T_M-1}\|_2\le1,\\
&\quad x(0)=0
\end{align*}
Here, the set $\mathcal{B}$ quantifies the maximum possible activation of each brain region under bounded input energy.
To compute $\mathcal{B}$, since $x(T_M)$ is a vector, one can optimize each of its entries individually (corresponding to specific brain regions). Given that the system dynamics $f(\cdot)$ are linear and the constraints are convex, this optimization can be efficiently solved using linear programming.
\begin{figure}[t]
  \centering
  \includegraphics[width=.9\linewidth]{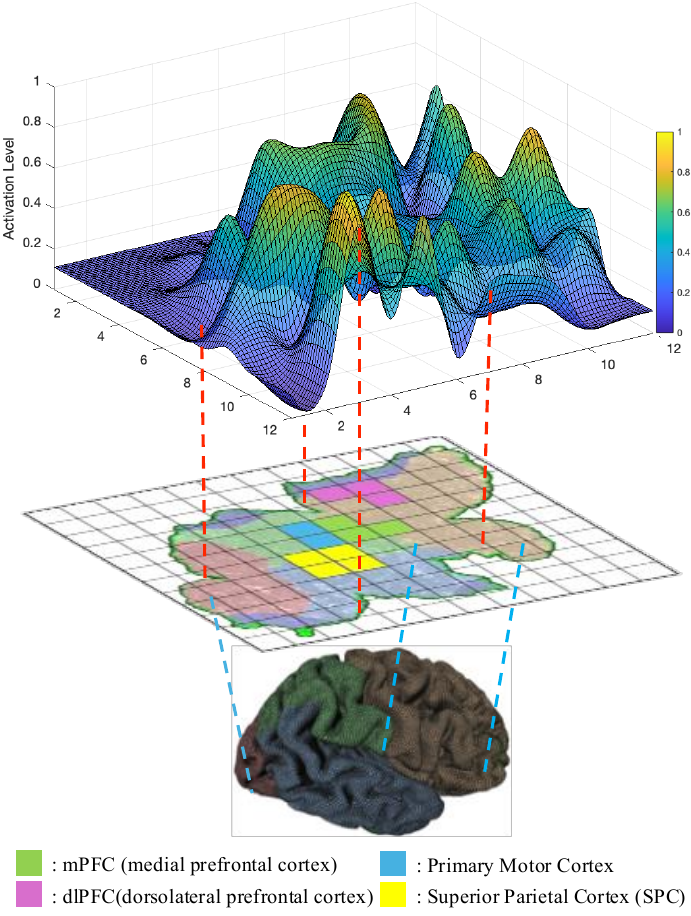}
  \caption{Three-tier illustration mapping the reachable-set activation surface (top) onto a flattened 12\,$\times$\,12 cortical grid (middle) and, from there, onto its anatomical locations in a 3-D brain rendering (bottom).  Colored tiles mark four representative regions.}
  \label{fig:activation_link}
\end{figure}

To visualize $\mathcal{B}$, we map its values onto a two-dimensional grid corresponding to the cortical parcellations (using a 90‐region atlas arranged in a 12×12 layout;  see Fig.~\ref{fig:activation_link}). Each region is colored according to its normalized reachability value, producing a \textit{reachability-based heatmap}. 
An example of this visualization, using real-world fMRI data, is shown in Fig.~\ref{fig:oneSubjectEightTasks}, Sec. ~\ref{sec:Reachability_vis}. Such heatmaps can be generated for each subject and each task condition to reveal the distinct spatial patterns of maximal reachability across brain regions. This offers a new way of brain network visualization, complementing standard correlation-based or connectivity-focused analyses that emphasize inter-regional interaction patterns.

\section{Results}
\label{sec:Results}
This section uses real-world datasets to verify the effectiveness of the proposed causal-based fingerprints for both subject and task fingerprints, as well as the reachability visualization.
\begin{figure}[ht]
\centering
\includegraphics[width=.45\textwidth]
{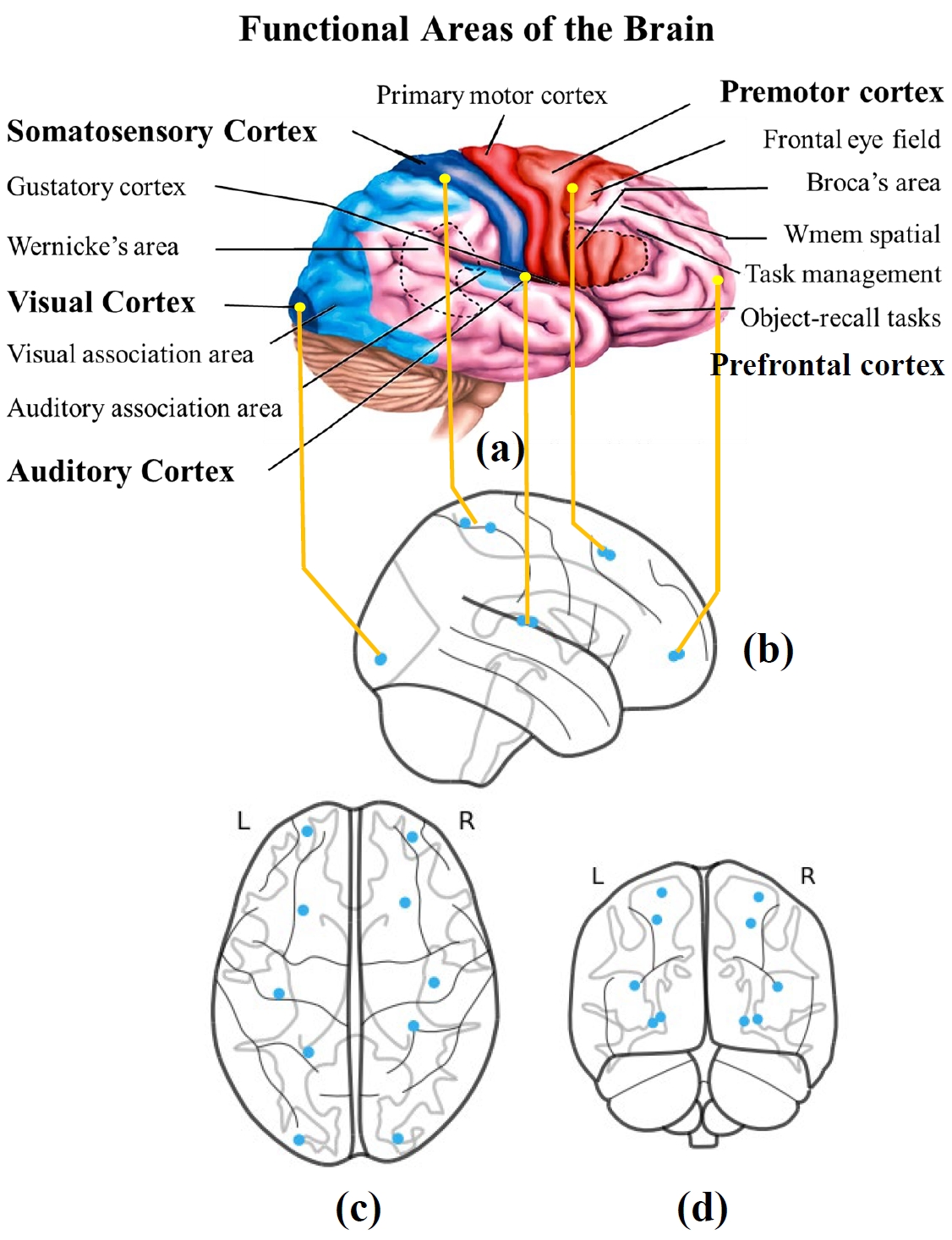}
\caption{(\textbf{a}) Functional Areas of the Brain. (\textbf{b-d}) Ten Schaefer Parcellation Areas (blue dots) that are Chosen as System Inputs: Two areas are associated with the prefrontal cortex; Two areas are associated with the pre-motor cortex; Two areas are associated with the somatic cortex; Two areas are associated with the vision cortex; Two areas are associated with the hearing cortex. We assume these are areas likely to `initiate' certain brain activities. \vspace{-1em}}
\label{fig:brain_map}
\end{figure}


\begin{table*}[ht]
\centering
\caption{Subject Fingerprinting Using Multiple Approaches}
\label{tab:my_label}
\setlength{\tabcolsep}{1pt} 
\renewcommand{\arraystretch}{1.05} 
\scriptsize   
\begin{tabular}{|l|c|c|c|c|c|c|c|c|c|c|}
\hline
\textbf{Method} & CM+MD\&P & CM+CoR & CM+FN & CM+GNN & FC+CoR & FC+MD\&P & FC+FN & FC+GNN & TSF & PH-H0 \\
\hline
Rest1 LR & 95.396\% & 58.397\% & 42.251\% & $<\!5\%$ & 45.439\% & 69.736\% & 33.952\% & $<\!5\%$ & 88.7\% & 85.6\% \\
Rest2 LR & 96.334\% & 61.295\% & 47.175\% & $<\!5\%$ & 45.436\% & 70.929\% & 37.574\% & $<\!5\%$ & 89.4\% & 86.2\% \\
Rest1 RL & 94.970\% & 57.118\% & 41.333\% & $<\!5\%$ & 41.603\% & 67.945\% & 38.698\% & $<\!5\%$ & 87.9\% & 84.7\% \\
Rest2 RL & 94.203\% & 58.978\% & 43.463\% & $<\!5\%$ & 46.803\% & 67.775\% & 36.737\% & $<\!5\%$ & 88.5\% & 84.3\% \\
\hline

\end{tabular}

\vspace{0.5em}
\begin{tabular}{@{}lll@{}}
CM: Causal Dynamics Model & FC: Functional Connectivity & CoR: Correlation \\
MD\&P: Modal Decomposition and Projection & GNN: Graph Neural Network & FN: Frobenius Norm \\
PH-H0: Persistent Homology(H0 Landscape)~\cite{WY-XJ-CY-YY-2025}  & TSF: Tangent-Space Fingerprinting~\cite{CD-KV-DJ-2025} &
\end{tabular}
\end{table*}

\begin{table}[ht]
\centering
\caption{Subject Fingerprinting Accuracy (\%) Comparing Single- vs.\ Two-Timescale Approaches.}
\label{tab:new_fingerprint}
\setlength{\tabcolsep}{1.5pt} 
\begin{tabular}{|l@{\hskip 0.1cm}|c@{\hskip 0.1cm}|c@{\hskip 0.1cm}|c|}
\hline
\raisebox{0.3ex}{\shortstack{\textbf{Method} \\ ~}} &
\shortstack{{\small One} \\ {\small Timescale}} & 
\shortstack{{\small Two Timescale} \\ {\small Fast+Slow}} & 
\shortstack{{\small Two Timescale} \\ {\small Slow only}} \\
\hline
Rest1 LR & 31.542\% & 81.245\% & 95.396\% \\
Rest2 LR & 30.937\% & 80.904\% & 96.334\% \\
Rest1 RL & 29.886\% & 78.517\% & 94.970\% \\
Rest2 RL & 30.123\% & 79.710\% & 94.203\% \\
\hline
\end{tabular}
\end{table}

\subsection{Dataset and Modeling} 
\label{sec:Preprocessing}
We utilize the Human Connectome Project (HCP) dataset~\cite{van2013wu}, which provides fMRI time-series data for 391 unrelated subjects. Each subject participated in two resting-state fMRI sessions and seven distinct task-based fMRI sessions: emotional response (Emot), gambling (Gamb), language processing (Lang), motor function (Moto), relational processing (Rela), social cognition (Soci), and working memory (WMem). Each subject-task pair has two recordings, scanned in LR (left to right) or RL (right to left) patterns. All fMRI data were parcellated using the Schaefer-100 atlas~\cite{schaefer2018local}, which partitions the cerebral cortex into $p = 100$ distinct regions. The fMRI time series for each region (parcel) was recorded at 720 ms intervals. In line with Remark 1, although higher-resolution parcellations are available, the 100-region resolution used here is sufficient for our purposes. This level of granularity allows us to investigate how the brain’s causal signatures can distinguish individual subjects and identify specific fMRI tasks, and our results confirm that this resolution is adequate for capturing those distinguishing signatures. 
In accordance with our model (Equation~\eqref{eq_model}), we select $n = 10$ out of the 100 brain regions as input nodes. These selected regions are located in the brain’s premotor and sensory cortices (see Fig.~\ref{fig:brain_map}), which is consistent with known functional hubs identified in~\cite{hardwick2015multimodal}. Specifically, the input set includes two parcels from each of the following cortical areas: prefrontal cortex, premotor cortex, somatosensory cortex, visual cortex, and auditory cortex. They were chosen due to their pivotal roles in various cognitive and sensory functions, including high-level task management, memory recall, spatial processing, somatosensory processing, visual perception, and auditory processing~\cite{hardwick2015multimodal}. The remaining $m = 90$ brain regions are treated as the system’s state in our model.

Specifically, we use the minimally pre-processed resting-state runs released by the UCSD Library~\cite{TAK:22:data,TU-AK-TE-AE-SL-2020} based on the HCP Young Adult raw data~\cite{rawdataHPC}.  
These scans have already undergone the standard HCP pipeline~\cite{GM-SS-WJ-CT-FB-2013}, which includes gradient-distortion and motion correction, EPI distortion correction, alignment to MNI space, high-pass filtering with a 2000-s cut-off, and ICA-FIX denoising~\cite{SK-GH-DG-BC-GM-2014}.  
After parcellation with the Schaefer-100 atlas, each run yields a $100\times1190$ matrix of regional time-series sampled every $0.72$ s~\cite{GM-SS-MD-AJ-AE-2016}.

\begin{figure*}[ht]
\centering
\includegraphics[width=\textwidth]{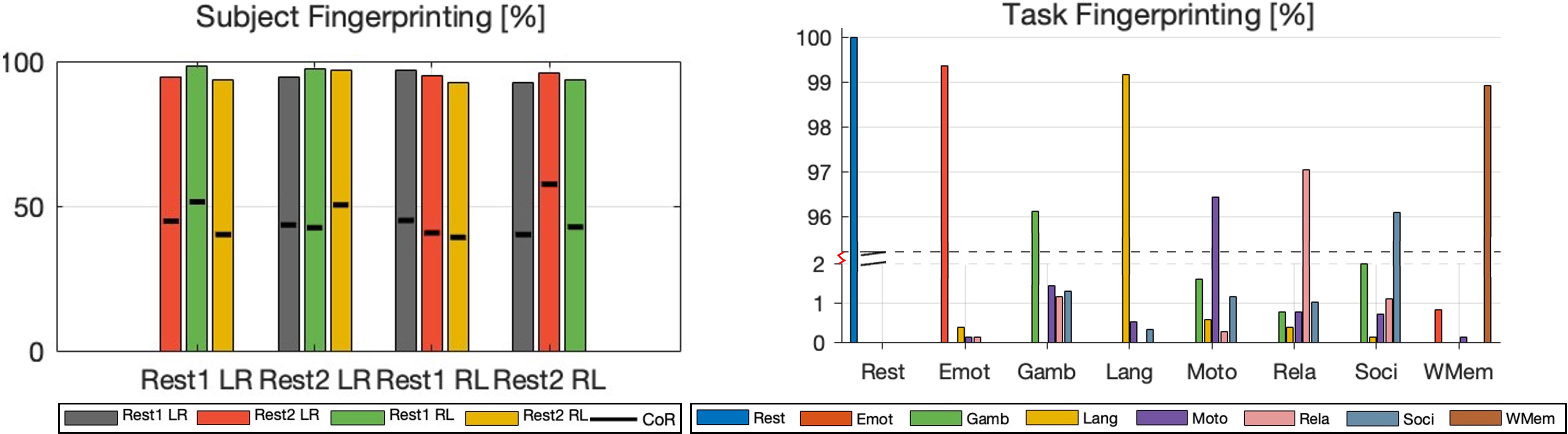}
\caption{Left: Subject Fingerprinting based on CM+MD\&P: The identification accuracy is compared with (black line) using the FC+CoR method.
Right: Task Fingerprinting based on CM+GNN: The figure shows the correct/incorrect classification for the same/different fMRI tasks. The y-axis is split. The top segment (95--100\%) shows the near-perfect accuracies, while the bottom segment (0--2\%) zooms in on the small error rates.
}
\label{fig:subject_task-accuracy}
\end{figure*}

\subsection{Subject Causal Fingerprinting}
We first verify the effectiveness of the proposed Causal-Dynamics Model combined with Modal Decomposition and Projection (CM+MD\&P) for subject fingerprinting. Previous research has established that resting-state fMRI shows considerable differences in brain activation patterns among individuals~\cite{raichle2006brain}. The dataset consists of two resting sessions, Rest1 and Rest2, each with two scanning orders, LR and RL. 
Preprocessing follows Section~\ref{sec:Preprocessing}. 

We run four cross-condition folds: each of the four resting scans (Rest1-LR, Rest1-RL, Rest2-LR, Rest2-RL) is taken in turn as the reference database, denoted as $\mathcal{R}_h$, while the remaining three scans serve as queries.  This guarantees that reference and query data are always disjoint. Within each fold, model parameters $(Q,A,B_1,B_2)$ are estimated only from the reference run; the ensuing modal matrices serve as one-shot “causal signatures” and are not updated during querying. Both baselines and the proposed method use the same four‐fold split and identical preprocessing.

The subject fingerprint is obtained by solving~\eqref{eq_dist}. For the results presented below, we chose the feature set as  $\mathcal{T}= [\widehat{T}]$, which relies only on the dynamic models of slow interaction and ignores the fast one. More details about this will be discussed in Remark~\ref{RM_2}.
The subject fingerprint result, presented in Table~\ref{tab:my_label}, demonstrate that the causal fingerprinting method, specifically CM+MD\&P, consistently achieves an accuracy rate above 94\% for all resting sessions. The highest accuracy observed is 96.334\% for the Rest2 LR session. This performance significantly surpasses that of the correlation-based baseline method, FC+CoR, which averages approximately 45\% accuracy. Even though capturing causal interactions is harder, our method not only outperforms the FC+CoR baseline but also other alternative approaches such as CM+CoR, CM+FN, and FC+MD\&P. Tangent-Space Fingerprinting (TSF) achieves $88\%$ accuracy, about $6$\, percentage points behind CM+MD\&P, because it models only the Riemannian geometry of static functional connectivity.  
The PH-H0 landscape method attains $85\%$; its zero-dimensional topological summary likewise ignores directionality and temporal scale, leaving it $10$\, percentage points short of our causal approach. To ensure a fair comparison, we applied the CM+MD\&P method as described in Sec.~\ref{sec_subjetid} and the FC+CoR method as outlined in the 2015 study by FES et al.~\cite{FES-SX-SD-RMD-HJ:15} directly to our dataset, without any further adjustments or preprocessing. Although additional data processing could potentially improve the accuracy of all methods tested, our main goal is to confirm the viability and efficacy of causal signatures for individual identification. Therefore, we have deliberately avoided such optimizations to maintain an impartial comparison.
To assess the generalizability of our approach, we ran experiments on a public dataset~\cite{MD-dM-GT-MM-SL-2015}. This dataset contains resting‑state fMRI scans from healthy controls and Alzheimer's disease patients. It does not include repeated sessions that match our original design. To work around this, we split one long resting‑state scan into two separate segments. This produced two distinct datasets for analysis. Our method achieved nearly 99\% accuracy in subject fingerprinting. These results highlight its strong performance and replicability across different populations and scanning protocols.

A detailed comparison is presented in Table~\ref{tab:my_label}, showing the average accuracy (across cases from Fig.~\ref{fig:subject_task-accuracy}-Left) for combinations of feature extraction methods (CM and FC) and classification methods (MD\&P, CoR, FN, and GNN). Here, FN refers to the Frobenius norm $\|R^Q - R^D\|_F$, used to measure distances between query and database matrices for classification. The proposed CM+MD\&P method achieves the highest performance among all combinations. Comparing FC+CoR and FC+MD\&P suggests that MD\&P can effectively evaluate similarities between FC matrices, leading to improved fingerprinting accuracy. The FN method is less robust than MD\&P for classification. And the GNN method is generally unsuitable for subject fingerprinting due to limited data availability. We also tested other deep learning models with convolution or attention mechanisms and various layer designs, but they similarly showed poor performance.

\begin{remark}\label{RM_2}[\textbf{\textit{Single vs. Two-Timescale Modeling and Feature selection}}]
To implement~\eqref{eq_dist}, the above result considers a two-timescale model, but uses only the slow dynamical modes, i.e., $(A, B_2)$, for subject fingerprinting. In addition to this configuration, one can also consider:
(i) using a single-timescale state-space model, or
(ii) using the full two-timescale dynamics, i.e., $(Q, A, B_1, B_2)$. 
In Table~\ref{tab:new_fingerprint}, we report the subject fingerprint accuracy achieved under these three configurations.
The single-timescale model achieves approximately 30\% accuracy, which indicates limited discriminative power when ignoring the multi-timescale nature of brain activity. 
Incorporating the full two-timescale dynamics significantly improves performance to around 80\% (presented in our prior work~\cite{SDC-SL-DTD-WX-2024}), which demonstrates the value of separating fast and slow modes. 
Surprisingly, the best results are achieved using only the reduced model $(A, B_2)$, which exceeds 94\% accuracy across various resting-state permutations. 
This result suggests that while accounting for both fast and slow dynamics is necessary for enhancing performance, the most distinctive subject-specific information resides in the slower portion of these dynamics. A potential reason might be that fMRI data are typically dominated by hemodynamic responses spanning seconds, the faster sub-interval interactions and the simultaneous input effects tend to be noisier and harder to resolve.
\end{remark}

\begin{table*}[ht]
\centering
\caption{Task Fingerprinting Accuracy Using Multiple Approaches}
\label{tab:my_label2}
\setlength{\tabcolsep}{0.75pt} 
\scriptsize 
{%
\begin{tabular}{|l|c|c|c|c|c|c|c|c|c|}
\hline
\textbf{Method} & \textbf{CM+GNN} & {CM+RF} & {CM+SVM} & {FC+GNN} & {R+GNN}& {R+DNN} & {R+DBRNN} & {R+BAnD}  & {R+TCN--BiLSTM} \\
\hline
Rest & \textbf{100}\% & \textbf{100}\% & 99.232\%  & 97.307\% & 96.629\% & 95.164\% & 94.355\% & \textbf{100}\% & 99.532\% \\
Emot & \textbf{99.872}\% & 99.488\% & 85.677\% & 89.160\% & 69.238\% & 72.912\% & 84.757\% & 98.493\% & 93.445\% \\
Gamb & \textbf{96.122}\% & 74.168\% & 93.094\% & 75.553\% & 81.683\% & 68.535\% & 69.023\% & 94.176\% & 89.395\% \\
Lang & \textbf{99.159}\% & 96.675\% & 89.514\% & 85.755\% & 89.177\% & 81.517\% & 82.204\% & 97.491\% & 91.304\% \\
Moto & \textbf{96.733}\% & 80.562\% & 96.675\% & 77.982\% & 76.858\% & 66.374\% & 77.773\% & 95.754\% & 88.992\% \\
Rela & \textbf{98.051}\% & 81.419\% & 81.841\% & 90.609\% & 86.554\% & 87.700\% & 79.628\% & 93.121\% & 89.663\% \\
Soci & \textbf{97.544}\% & 75.703\% & 77.749\% & 69.811\% & 73.706\% & 71.455\% & 69.950\% & 96.385\% & 92.118\% \\
WMem & \textbf{99.523}\% & 90.767\% & 95.396\% & 78.760\% & 80.202\% & 76.620\% & 82.446\% & 98.299\% & 94.327\% \\
\hline
{Average} & \textbf{98.376}\% & 87.350\% & 89.897\% & 83.117\% & 81.756\% & 78.929\% & 80.017\% & 96.710\%  & 92.347\% \\
\hline
\end{tabular}
}
\begin{tabular}{@{}lll@{}}
CM: Causal Dynamics Model & GNN: Graph Neural Network & DNN: Deep Neural Network~\cite{WX-LX-JZ-NB-ZY-2020} \\
FC: Functional Connectivity & SVM: Support Vector Machines & BAnD: Brain Attend and Decode~\cite{nguyen2020attend} \\
R: Raw data & \hspace*{-1.8cm}RF: Random Forest & \hspace*{-3cm}DBRNN: Deep Bidirectional Recurrent Neural Network~\cite{ZH-SH-FA-LO-WU:2022} 
\end{tabular}
TCN--BiLSTM: Temporal Convolutional Network--Bidirectional Long Short--Term Memory~\cite{KM-GJ-BJ-BK-2025}

\end{table*}

\subsection{fMRI Task Causal Fingerprinting.}
We validate the effectiveness of the proposed Causal Modeling (CM) combined with Graph Neural Network (GNN) method for task fingerprinting, as presented in Table~\ref{tab:my_label2}. Using causal signatures from the HCP dataset, we split the data evenly, with 50\% allocated for training and 50\% for testing. The model achieves perfect accuracy (100\%) for Rest and near-perfect accuracy for Emot, Lang, WMem. However, tasks such Gamb, Moto, Rela, and Soci exhibit slightly lower performance, with most misclassifications occurring among these tasks.

To illustrate the strengths of our CM+GNN method, we compared it to several state-of-the-art approaches across eight tasks~\cite{nguyen2020attend,LA-ST-ST-ST-CH:2005,ZH-SH-FA-LO-WU:2022}\footnote{For non-open-source methods, we reproduced the results based on the descriptions provided in the original papers.}. 
Other columns in Table~\ref{tab:my_label2} report the average accuracy of these methods. Our CM+GNN method consistently outperforms others, achieving the highest average accuracy (98.376\%), which highlights its robustness in task fingerprinting. For Emot and Lang, CM+GNN effectively identifies unique neural patterns, surpassing CM+RF and Raw+DBRNN. Performance decreases slightly for tasks like Gamb, Moto, and Soci, possibly due to overlapping neural activity. Similar declines are observed in other methods, such as Raw+DNN and Raw+DBRNN, for these challenging tasks.

We also evaluated the R + TCN--BiLSTM baseline, a Temporal Convolutional Network followed by bidirectional LSTM layers trained end to end on raw fMRI series~\cite{KM-GJ-BJ-BK-2025}. 
Its mean accuracy of $92.3\%$ is higher than the other raw‑data deep nets (R + GNN, R + DNN, R + DBRNN) yet remains about $6$ percentage points below CM + GNN. While the convolutional blocks capture local temporal motifs, the model lacks directed and multi‑timescale information, limiting its discriminative power relative to our causal signatures.

To further assess the reliability of our causality-based approach, we tested it on an independent dataset~\cite{PI-AL-AM-AL-GA-2020}. The method achieved nearly 95\% accuracy in task fingerprinting, confirming its consistency and wide applicability.

\subsection{Effect of sampling interval $\Delta t$ on model fingerprinting accuracy}
\begin{table}[ht]
\centering
\caption{Subject fingerprinting accuracy (\%) under different sampling intervals $\Delta t$ (seconds)}
\label{tab:delta_t_sensitivity}
\setlength{\tabcolsep}{3pt}
\begin{tabular}{|l|c|c|c|}
\hline
\textbf{$\Delta t$ (s)} & $\mathbf{\Delta t = 0.72}$ & $\mathbf{\Delta t = 1.44}$ & $\mathbf{\Delta t = 2.16}$\\
\hline
Rest1 LR & 95.396\% & 90.143\% & 80.636\% \\
Rest2 LR & 96.334\% & 92.002\% & 81.452\% \\
Rest1 RL & 94.970\% & 88.747\% & 78.890\% \\
Rest2 RL & 94.203\% & 89.563\% & 77.938\% \\
\hline
\end{tabular}
\end{table}
Note that the sampling interval $\Delta t$ is critical to our discrete-time causal model. In this experiment, we adopt the native sampling interval $\Delta t = 0.72$s because it is the scanner’s repetition time (TR), as specified in the data description~\cite{rawdataHPC}. Repetition time is the time interval between the beginning of one MRI pulse sequence and the beginning of the next for the same slice. Staying at this original resolution avoids any resampling artifacts and preserves the rapid hemodynamic changes our two-timescale model is meant to capture.

Nevertheless, it is still important to discuss the sensitivity of our algorithm to the sampling interval $\Delta t$. Table~\ref{tab:delta_t_sensitivity} summarizes how resampling affects subject fingerprinting.  Accuracy is computed for each of the four resting runs at three resolutions: the native $0.72$s, and down‑sampled versions at $1.44$s (every second frame) and $2.16$s (every third frame).

The results in Table~\ref{tab:delta_t_sensitivity} confirm that accuracy peaks at the native interval ($\Delta t = 0.72$s), the scanner’s inherent repetition time in the HCP protocol and therefore the finest temporal resolution available. Accuracy drops by roughly five percentage points when the data are down-sampled to $1.44$ s and falls a further ten points at $2.16$ s. The two-timescale model evidently relies on sub-second variations that are smoothed out at coarser resolutions, degrading the estimated matrices $Q \& A$ blocks. We therefore retain $\Delta t = 0.72$ s for all reported experiments to preserve temporal fidelity while avoiding unnecessary information loss.\\
Task fingerprinting is less sensitive: accuracy falls from $98.376\%$ at $\Delta t = 0.72$ s to $96.442\%$ at $1.44$ s and $92.127\%$ at $2.16$ s, suggesting that large-scale task-evoked activation patterns remain highly discriminative even when the temporal resolution is moderately reduced.

\section{Visualization: Directed Causal Interaction and Reachability Landscape} 
\label{sec:ReachabilityVisualization}
\begin{figure}[h]
    \centering
    \includegraphics[width=.45\textwidth]{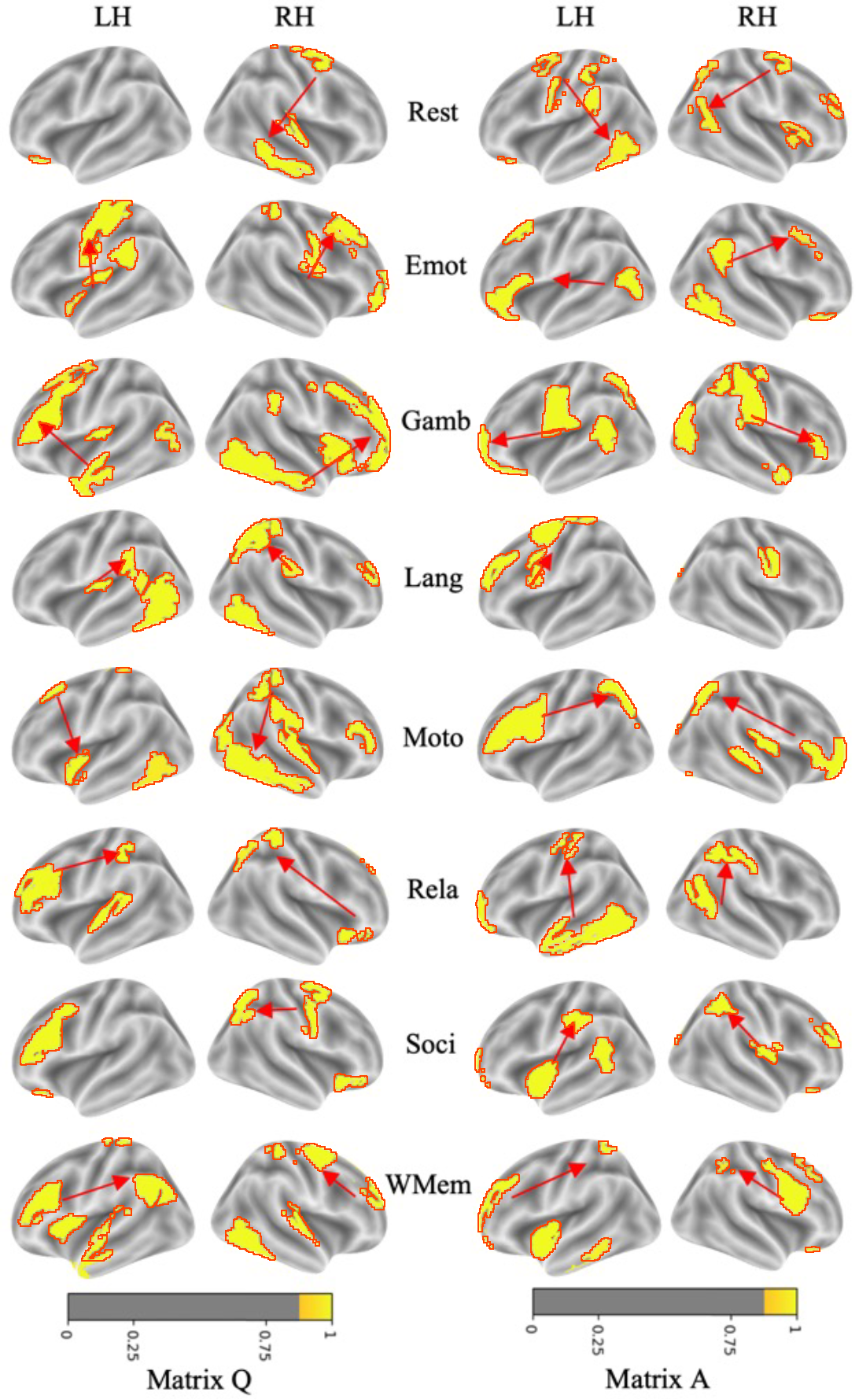}
    \caption{Visualization of Brain Regions with Strongest Connection Weights, Showing Association Strengths Across Multiple Cognitive Tasks (Importance Scores Indicated by Color-Bar). LH refers to the left hemisphere, and RH refers to the right hemisphere of the brain. Arrows indicate directed relationships, illustrating which regions exert causal influence on others. For clarity of presentation, we employ a two-color scheme and apply a threshold to highlight only the most active brain areas and their directional interactions.}
    \label{fig:gnn_brain_regions}
\end{figure}

\subsection{Directed Causal Interaction}
We use Fig.~\ref{fig:gnn_brain_regions} to present the causal signatures extracted for the eight HCP tasks (Rest, Emot, Gamb, Lang, Moto, Rela, Soci, and WMem) and serve as the foundation for our functional interpretation. Before plotting, the matrices $Q$ and $A$ are rescaled to a same range, which enables cross-task comparison on a unified metric. Each entry of $Q$ or $A$ conveys both magnitude and sign, which reveals the direction and excitation/inhibition relations of causal drive between pairs of regions. We use arrows in the figures to indicate the directions of these causal influences, which are discussed in detail for each task in the subsequent analysis. 
By separating matrix $Q$ and matrix $A$, we can observe how brain activity evolves over different timescales. Matrix $Q$ captures fast synchronous interactions among brain regions, reflecting real-time activation patterns during specific tasks. In contrast, Matrix $A$ illustrates slower dynamic evolutions, showing how current states are influenced by previous states. 

For the resting state data, brain interactions concentrate within the default mode network (DMN), especially the medial prefrontal cortex (mPFC) and posterior cingulate cortex (PCC), which support introspective thinking, self-related processing, memory integration and rest-task dichotomy~\cite{duong2021morphospace}. The mPFC exerts influence over the PCC, modulating introspective and self-related processing.
The emotion task shows strong connections in the amygdala, prefrontal cortex, and anterior insula. This is due to their roles in emotion generation and regulation. 
The gambling task primarily shows strong connections in the ventral striatum, prefrontal cortex, and insula, which are crucial for reward prediction and decision-making. The ventral striatum influences the prefrontal cortex and insula, affecting decision-making and reward processing.
The language task primarily shows strong activations in Broca’s area (Brodmann areas 44 and 45~\cite{DA-AR-DA-HA:1992}) and Wernicke’s area (Brodmann area 22) in the left hemisphere. Broca’s area is crucial for language production, while Wernicke’s area is essential for language comprehension. Broca’s area influences Wernicke’s area, facilitating coordinated language processing.
Motor execution engages the primary motor cortex, supplementary motor area (SMA), and basal ganglia. This reflects their active roles in movement planning and control. Causal arrows show the primary motor cortex directing the SMA as actions unfold.
For the relational reasoning task, the dorsolateral prefrontal cortex (DLPFC) and posterior parietal cortex provide the principal scaffold for working memory, abstraction, and relational encoding. The DLPFC leads the parietal cortex to facilitate complex cognitive manipulation.
Social cognition involves strong coupling between the mPFC and superior temporal sulcus (STS). These are regions implicated in theory of mind and socio-emotional appraisal. The mPFC drives the STS to modulate social evaluative processing.
During working memory maintenance, robust edges connect the DLPFC, parietal cortex, and anterior cingulate cortex (ACC). This aligns with their critical roles in information storage and monitoring. Here, the DLPFC influences parietal regions to sustain mnemonic content.

It can be observed that certain tasks share overlapping strong connections in specific regions. For example, the gambling, social, and emotional tasks all rely on pathways linking the prefrontal cortex and insula, caused by the collaborative role of these regions in risk-related decision making, emotion processing, and social cognition. Language and relational tasks converge within left temporal structures, notably Wernicke’s area and the hippocampus, which highlights their joint involvement in linguistic analysis and declarative memory. Motor and working memory conditions depend on coordinated activity between the DLPFC and parietal cortex, underscoring these regions’ shared contribution to motor sequencing and the active maintenance of information. This might be one of the reasons for the performance degradation (commonly shared by most methods) for task fingerprinting.

\begin{figure}[ht]
\centering 
\includegraphics[width=0.35\textwidth]{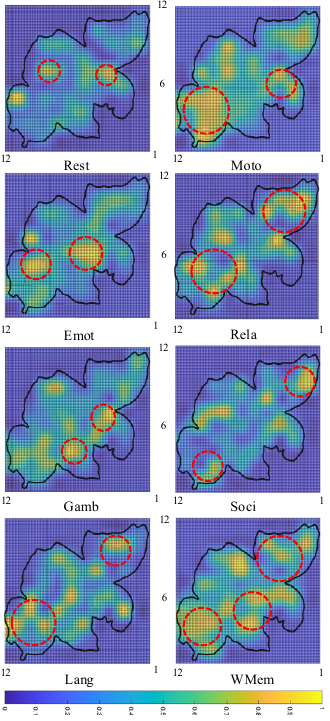} \caption{Reachability-based heatmaps for one subject across eight cognitive tasks (Rest, Emot, Gamb, Lang, Moto, Rela, Soci, and WMem). Brighter colors (yellow) indicate stronger maximum neural activations, while darker tones (blue) reflect lower excitability. Red dashed circles highlight representative activation hotspots corresponding to critical brain regions known for each cognitive task. The black boundary outlines the edge of the flattened cortical map. The correspondence between this reachability representation and anatomical brain regions is illustrated in Fig.~\ref{fig:activation_link}. These patterns highlight functional specialization differences across diverse cognitive tasks.} \label{fig:oneSubjectEightTasks} 
\end{figure}
\subsection{Reachability Landscape}
\label{sec:Reachability_vis}
Fig.~\ref{fig:oneSubjectEightTasks} illustrates reachability-based heatmaps describing the potentially highest activation level for a single subject across eight cognitive modes: Rest, Emot, Gamb, Lang, Moto, Rela, Soci, and WMem. Each heatmap represents normalized activation levels computed from a 90-region cortical parcellation arranged on a 12×12 spatial grid. Brighter areas (yellow) indicate higher neural excitability under bounded input conditions, whereas darker regions (blue) denote lower activation potentials.
During the resting state (Rest), the default‑mode network (DMN) shows the strongest activation. The medial prefrontal cortex (mPFC) and posterior cingulate cortex (PCC) dominate this pattern~\cite{RA-MA-2015}. 
During emotional processing (Emot), the limbic system becomes highly active. The ventromedial prefrontal cortex (vmPFC), amygdala, and insula lead this response~\cite{PK-WA-TD-TA-SF-2004}.  
In the gambling task (Gamb), the orbitofrontal cortex and ventral striatum show clear clusters of activation~\cite{KB-AD-CM-FO-GW-2001}. These clusters signal heightened responsiveness linked to reward anticipation and decision‑making. 
Language-related activations (Lang) appear primarily within classical left-lateralized perisylvian areas, prominently involving Broca’s area (inferior frontal gyrus) and Wernicke’s area (superior temporal gyrus)~\cite{FR-AD-BR-JE-LO-2011}. 
Motor execution (Moto) triggers the motor network. The primary motor cortex, supplementary motor area, and nearby premotor regions show strong excitation~\cite{LO-MA-HA-UL-2006}. 
Relational reasoning (Rela) engages the frontoparietal control network. The dorsolateral prefrontal cortex (dlPFC) and posterior parietal cortex display distinct activation foci~\cite{FE-EV-2014}. 
Social cognition (Soci) relies on theory‑of‑mind regions. The medial prefrontal cortex and temporoparietal junction (TPJ) dominate the pattern~\cite{SA-RE-BC-SI-2006}. 
Working memory (WMem) activates a frontoparietal network. The dorsolateral prefrontal cortex and superior parietal cortex show synchronized engagement~\cite{CU-CE-DE-MA-2003}. 
These reachability‑based visualizations provide individualized activation maps. They validate established neurofunctional models across cognitive domains. They also demonstrate the capability of reachability analysis for revealing concentrated brain regions under different modes.

\section{Conclusion}

To the best of our knowledge, we are among the first to investigate and quantify large-scale brain fingerprints in the context of causal dynamics.
This paper proposed new methods that employ an implicit-explicit discretization scheme to derive a new state-space model, which captures causal signatures as directed interactions with two-timescale temporal resolution. 
The causal dynamics model also enabled a novel visualization tool: the reachability landscape of brain states, which quantitatively represents the possible activation levels that the brain regions can reach under various fMRI tasks. The effectiveness of the proposed methods and visualization tools was validated using real-world datasets and through comparisons with other state-of-the-art techniques.

Despite the novelty of the proposed approach, we remark on the following limitations:
i) In our approach, system inputs were manually specifically selected based on different functional brain regions, while other ways of choosing system state input regions may potentially further improve the performance. 
ii) While our subject fingerprinting method is model-based and generalizable without training, the task fingerprinting method relies on training a Graph Neural Network (GNN), which may limit its generalization across datasets.
iii) Subject fingerprinting shows measurable sensitivity to the sampling interval $\Delta t$ (coarser sampling degrades accuracy).

Regarding clinical applications, our proposed method has the potential for the development of diagnostic and prognostic tools for neurodegenerative diseases such as Alzheimer's Disease (AD). By analyzing functional and directional influences between brain regions, the proposed approach can associate changes in brain network dynamics with the emergence of clinical symptoms.

In future work, we plan to:
i) Explore causal fingerprints and the reachability landscape concept in both healthy controls (e.g., through other open neuroimaging datasets) and in cases of neurodegenerative diseases such as Alzheimer's disease~\cite{xu2021optimization,xu2022consistency,garai2024effect};
ii) Enhance the generalizability of our task fingerprinting approach, potentially through transfer learning techniques to accelerate adaptation to new datasets.
iii) Systematically assess robustness to choices of sampling interval $\Delta t$, parcellation, and preprocessing, and report performance in a manner that is independent of the acquisition and processing pipeline;
iv) Extend the two-timescale linear model to piecewise linear or mildly nonlinear dynamics with uncertainty quantification, and replace manual selection of input regions with a data-driven procedure to reduce design bias.

\section{Acknowledgement}
This work was supported in part by the NSF ECCS Award 2332210, 4-VA, a collaborative partnership for advancing the Commonwealth of Virginia, NIH Grants U01 AG068057 and RF1 AG068191.


\section*{Appendix}
\subsection*{Details of GNN Training}

\begin{table*}[ht]
\centering
\caption{Paired $t$-tests on per-subject accuracy differences ($n=391$). 
$\Delta$Acc = (proposed $-$ baseline) in percentage points; Bonferroni thresholds: 
$\alpha_{\text{subj}}=0.0071$ (7 contrasts), $\alpha_{\text{task}}=0.0063$ (8 contrasts).}
\label{tab:t_tests_join}
\renewcommand{\arraystretch}{1.12}
\setlength{\tabcolsep}{6pt}

\begin{minipage}{0.48\textwidth}
\centering
\textbf{Subject -- CM+MD\&P vs.\ baselines}\\[2pt]
\begin{tabular}{l S[table-format=+2.1] S[table-format=1.1e-1]}
\toprule
Baseline & {\(\Delta\)Acc [pp]} & {$p$-value} \\
\midrule
CM+CoR    & +36.2 & 4.7e-22 \\
CM+FN     & +51.6 & 8.2e-27 \\
FC+CoR    & +48.8 & 2.9e-24 \\
FC+MD\&P  & +24.7 & 2.4e-14 \\
FC+FN     & +57.5 & 1.0e-33 \\
TSF       &  +6.6 & 2.1e-07 \\
PH-H0     & +10.0 & 4.5e-09 \\
\bottomrule
\end{tabular}
\end{minipage}
\hfill
\begin{minipage}{0.48\textwidth}
\centering
\textbf{Task -- CM+GNN vs.\ baselines}\\[2pt]
\begin{tabular}{l S[table-format=+2.1] S[table-format=1.1e-1]}
\toprule
Baseline & {\(\Delta\)Acc [pp]} & {$p$-value} \\
\midrule
CM+RF         & +10.4 & 1.1e-12 \\
CM+SVM        &  +8.1 & 4.2e-11 \\
FC+GNN        & +14.6 & 3.6e-12 \\
R+GNN         & +15.8 & 6.8e-13 \\
R+DNN         & +18.7 & 4.8e-14 \\
R+DBRNN       & +17.5 & 9.9e-14 \\
R+BAnD        &  +1.5 & 1.6e-02 \\
R+TCN--BiLSTM &  +6.0 & 3.2e-10 \\
\bottomrule
\end{tabular}
\end{minipage}
\end{table*}

We train the GNN in a supervised fashion using all $R_{s,h}^{D}$ labeled by task $h$.  Each node in the graph corresponds to one brain region, whose feature vector is that region’s row in $R_{s,h}$; the normalised $A$ block sets up the adjacency.  We minimise a negative log-likelihood loss comparing the GNN’s predicted task label to the true $h$. To enhance generalisation and prevent over-fitting, we employ weight decay (L2 regularisation) within AdamW, in tandem with SAM~\cite{FT-PE-KR-AL-MI-2020} as a meta-optimiser.  Specifically, in each training iteration SAM and AdamW operate in concert: AdamW (learning rate $1\times10^{-3}$, weight decay $5\times10^{-4}$) performs the parameter update, while SAM adds a perturbation step of radius $\rho = 0.02$ to avoid sharp minima.  The learning rate is reset every $T_{0}=20$ epochs by a cosine warm-restart schedule; each cycle length doubles via $T_{\mathrm{mult}}=2$ and decays to $\eta_{\min}=10^{-6}$.  We use mixed precision (AMP), gradient clipping at $1.0$, mini-batches of size $32$, and early stopping with patience $20$. On-the-fly data augmentation further improves robustness: Gaussian feature noise with $\sigma = 0.1$, random edge drop of $20\%$, and subgraph sampling that keeps $80\%$ of the nodes. At inference time, a query $R^{Q}$ is fed into the same pipeline, producing a log-softmax distribution over tasks; the highest score is taken as the predicted label.  Through this causal-based GNN we leverage the multi-timescale structure of $\bigl[Q\,A\,B_{1}\,B_{2}\bigr]$ to classify fMRI tasks with high accuracy.
\subsection*{Results Validation}

For each baseline we construct a length-$391$ vector of subject‑wise accuracies (one paired value per subject, averaged over the four cross‑condition folds) and apply a two‑sided paired $t$‑test after verifying normality with the Shapiro–Wilk test.  Under a Bonferroni adjustment for the seven subject‑level comparisons ($\alpha = 0.05/7 \approx 0.0071$), every contrast with the proposed CM+MD\&P remains highly significant, with $p$‑values spanning $4.7\times10^{-22}$ to $<10^{-33}$ (Table~\ref{tab:t_tests_join}, left).

For task fingerprinting we test eight baselines, so the corrected threshold is $\alpha = 0.05/8 \approx 0.0063$.  All baselines, including the newly added R\,+\,TCN–BiLSTM, are significantly different from the proposed CM+GNN except R\,+\,BAnD, whose $p = 1.6\times10^{-2}$ exceeds the threshold (Table~\ref{tab:t_tests_join}, right).

Mean accuracies with $95\,\%$ confidence intervals across folds are $95.4\%\pm0.3$ for subject fingerprinting and $98.4\%\pm0.2$ for task fingerprinting, indicating that the estimates are stable.  Ten random perturbations of the reference/query split alter subject accuracy by no more than $0.6$ percentage points (standard deviation), confirming that the results are not driven by a particular split.

\bibliography{sn-bibliography}

\end{document}